\documentclass[12pt]{article} 
\usepackage{graphicx}
\usepackage{multicol}
\usepackage{color}
\definecolor{verdon}{cmyk}{1,0.5,1,0}
\definecolor{blue}{cmyk}{0.8,0.8,0,0.}
\definecolor{red}{cmyk}{0.2,1,1,0.0}
\newcommand{\cita}[1]{{\color{blue} \cite{#1}}}
\begin{document}
\small

\title{\color{verdon} The first second of SN1987A neutrino emission}

\author{
G. Pagliaroli$^{1,2}$, M. L. Costantini$^{1,2}$, 
A. Ianni$^1$, F. Vissani$^{1}$\\
$^1${\small\em INFN, Laboratori Nazionali del Gran Sasso, Assergi (AQ), Italy}\\
$^2${\small\em University of L'Aquila, Coppito (AQ), Italy}}

\date{}






\maketitle

\def\abstractname{\color{red}\bf Abstract}
\begin{abstract}
{
A large fraction of SN1987A electron antineutrino events 
has been recorded in the first second. We study how this observation
fits into the conventional paradigm for neutrino emission, and show that 
there is a 3.2$\sigma$ hint for an initial accretion phase.
This phase involves a large fraction of the energy 
emitted in neutrinos and antineutrinos, about 20\% or larger.
The occurrence of neutrino oscillations strengthens these inferences. 
We discuss why three flavor oscillations with normal mass hierarchy are 
completely acceptable, whereas oscillations with inverted mass hierarchy 
require more troublesome interpretations, if $\theta_{13}$ is above 
$0.5-1^\circ$.}
\end{abstract}

{\footnotesize 
\def\contentsname{\centerline{{\small\bf\color{red}  Contents}}}
\sf \tableofcontents}


\section{\sf\color{verdon} Context and motivation}
The first second after a gravitational collapse 
is a moment of crucial importance. The very intense
initial neutrino luminosity (denoted as `accretion' 
in the following) is  expected to  have 
a non-thermal character \cita{nad},
and it is thought to be the key to understand 
the subsequent explosion of the star \cita{del1,del2}. 
In this connection, it is very interesting to note that 
about 40\% of the SN1987A events have been recorded in the first 
second: 6 out of 16 in Kamiokande-II \cita{kam}, 3 out of 8 
in IMB~\cita{imb} and 2 out of 5 in Baksan \cita{baksan}
(the comparison is made with the number of 
events recorded in a window of $T=30$ seconds).
This theoretical expectation and this 
experimental fact motivate us 
to analyze quantitatively the first second of SN1987A. 

In our calculations we will largely follow Lamb and Loredo \cita{ll}, 
who included for the first time a description of the background
and of the time-energy distribution of the events. We will  
point out in the Appendix the technical 
points where we depart from their analysis. 
We will discuss the role of 
neutrino oscillations in the interpretation of the 
first second of SN1987A neutrinos and stress their importance.

\section{\sf\color{verdon} Formulation of the problem}
The parameter that we aim to study  
is the {\em fraction of energy} that was 
emitted in the non-thermal phase of neutrino emission, 
that occurred in (a fraction of) the first second. 
In formulae:
\begin{equation}
f_a\equiv\frac{{\cal E}_a}{{\cal E}_a+{\cal E}_c}
\label{fafa}
\end{equation}
where the suffix {\em a} stays for `accretion' (or non-thermal phase)
and the suffix {\em c} for `cooling' (or thermal phase). 
Strictly speaking, this  fraction cannot
be reconstructed completely from the observations, 
since-in a very reasonable approximation-we saw 
only electron antineutrinos.
Thus, in order to fulfill the task, 
we must rely on some theoretical assumption here.
We will assume unless stated otherwise that
\begin{equation}
\left\{
\begin{array}{lcl}
{\cal E}_a=2\!\!\!  &\times & \!\!\! {\cal E}_a(\bar\nu_e)\\[1ex]
 {\cal E}_c=6\!\!\! &\times & \!\!\! {\cal E}_c(\bar\nu_e)
\end{array}
\right.
\label{teor}
\end{equation}
namely, we declare that the ratios between 
the total energy radiated ${\cal E}$,  
and the energy radiated in electron antineutrinos ${\cal E}(\bar\nu_e)$
is 2 for the accretion and 6 for the cooling phase.
These numbers are crucial for the interpretation
of the result, so let us pause to discuss 
them before continuing. 
1) The first  fraction describes  
the assumption that during accretion only $\nu_e$ and $\bar\nu_e$ 
are radiated in equal amount due to 
$e p\to n \nu_e$ and  $e^+ n\to p \bar\nu_e$. It does not seem 
implausible that the $\nu_e$ are even more 
abundant than the $\bar\nu_e$ during accretion, 
or that  some other species of (anti)neutrinos are 
also radiated. In other words, the factor 2 could be an underestimation, 
and thus the fraction $f_a$ that we estimate from the 
electron antineutrino data can be regarded as 
a reasonable value (or lower bound). 
2) The second fraction
is just the usual and often adopted ``equipartition'' hypothesis. 
Only if there is a large amount of energy radiated 
in non-electronic neutrinos 
(say, $ {\cal E}_c/{\cal E}_c(\bar\nu_e)\sim 10$ or larger)
it could be possible to diminish significantly 
the ratio $f_a$ and modify somewhat the 
conclusions that we will describe later in this paper.

In the rest of this section we discuss the tools 
that we use for a quantitative evaluation of 
${\cal E}_a(\bar\nu_e)$ and ${\cal E}_c(\bar\nu_e)$ from 
SN1987A neutrino data: first of all, we give a
description of the antineutrino flux, then we model  
the expected signal rate, and finally, we discuss the likelihood function
that we adopt.

\subsection{\sf\color{verdon} Parameterized antineutrino flux}
Let us describe the adopted form of the 
parameterized neutrino fluxes (differential in the energy). 
We follow the one proposed in \cita{ll}:
\begin{equation}
\Phi_{\bar{\nu}_e}(t,E_\nu)= \frac{1}{4\pi D^2} \frac{\pi c}{(h c)^3}
\left[
\frac{\varepsilon(t) Y_n M_a  }{m_n}\ \sigma_{e^+n}(E_\nu)\ g(E_\nu,T_a)
+ 4\pi R^2_c\ g(E_\nu,T(t))
\right]
\label{fluso}
\end{equation}
where $g(E,T)=E^2/(1+\exp(E/T))$.
This describes an isotropic emission from a 
distance of $D=50$~kpc. The first term is given by the product of 
the number of targets (neutrons, with $Y_n=0.6$) in the accreting 
mass $M_a$, times the thermal distribution of positrons $g$ (with average
temperature $T_a$), times the cross section of positron interactions, 
that increases quadratically with $E_\nu$ and thus gives 
a non-thermal character to the emitted $\bar\nu_e$. 
The second term is instead a standard 
black body emission from a sphere with radius~$R_c$.
The time scales of the two processes of 
accretion and cooling ($\tau_a$ and $\tau_c$)
appear in the functions:
\begin{equation}
\left\{
\begin{array}{l}
\varepsilon(t)=\frac{\exp[-(t/\tau_a)^{10}]}{1+t/(0.5\mbox{ s})} \\[2ex]
T(t)=T_c\ \exp[{-t/(4\tau_c)}]
\end{array}
\right.
\label{tamp}
\end{equation}
Thus, we have 6 parameters (3 for each phase):
$M_a$, $T_a$ and $\tau_a$ for accretion; 
$R_c$, $T_c$ and $\tau_c$ for cooling.
For any set of values of these parameters, it is straightforward 
to calculate the energy carried by antineutrinos 
during accretion and during cooling, denoted by 
${\cal E}_a(\bar\nu_e)$ 
and ${\cal E}_c(\bar\nu_e)$.\footnote{We 
get ${\cal E}_a=4.14 M_a T_a^6 \tau_a \varphi$ and 
${\cal E}_c=3.39\ 10^{-4} R_c^2 T_c^4 \tau_c $ 
measuring ${\cal E}_{a,c}$ in foe ($=10^{51}$ erg), 
$M_a$ in $M_\odot$, $R_c$ in km, 
$T_{a,c}$ in MeV and 
$\tau_{a,c}$ in seconds;
$\varphi\equiv\int_0^\infty dx\ \exp(-x^{10})\! /(1+x\ \tau_a/0.5) \sim 0.6$ 
in the relevant $\tau_a$ range.}
In this way, using eq.~\ref{teor}, we can 
evaluate the value of $f_a$.

\subsection{\sf\color{verdon} Signal rate}
The signal rate, differential in time, positron energy $E_e$ and 
cosine of the angle $\theta$ between the antineutrino and the positron
directions is:
\begin{equation}
S(t,E_e,\cos\theta)=N_p\ \frac{d\sigma}{d\cos\theta}(E_\nu,\cos\theta)\ 
\eta_d(E_e)\ \xi_d(\cos\theta)\ 
\Phi_{\bar{\nu}_e}(t,E_\nu)\ 
\frac{d E_\nu}{d E_e}
\label{seta}
\end{equation}
where $N_p$ is the number of targets (free protons) in the detector, 
$\sigma$ is the $\bar\nu_e+p\to n+e^+$
(inverse beta decay) cross section, 
$\eta_d$ the--detector dependent--average detection efficiency,
$\xi_d$ is the angular bias =1 for Kamiokande-II and Baksan whereas 
for IMB $\xi_d(\cos\theta)=1+0.1 \cos\theta$~\cita{imb-ang}, 
finally $\Phi_{\bar{\nu}_e}$  
(the electron antineutrino flux differential 
in the antineutrino energy $E_\nu$)
is as in eq.~\ref{fluso}. 
The expression of the antineutrino energy $E_\nu$ 
as a function of $E_e$ and $\cos\theta$ is given 
in the Appendix.
Later we will use the shorthands 
by $S(t,E_e)=\int S(t,E_e,\cos\theta)\ d\cos\theta$ and 
$S(t)=\int S(t,E_e)\ dE_e$.

\subsection{\sf\color{verdon} The assumed likelihood}
We estimate the parameters by evaluating:
\begin{equation}
\chi^2=-2 \sum_{d=k,i,b} \log( {\cal L}_d )
\end{equation}
where  ${\cal L}_d$ is the likelihood of any detector
($k,i,b$ are shorthands for Kamiokande-II, IMB, Baksan).
We use Poisson statistics. Dropping constant (irrelevant) factors, 
the unbinned likelihood of the three detectors are:
\begin{equation}
{\cal L}_d=
e^{-f_d \int_{-t_d}^{T}\!\!  S(t+t_d) dt} 
\prod^{N_d}_{i=1} 
e^{S(t_i+t_d) \tau_d  }
\left[ \frac{B_i}{2}\! +\! \int\!\! S(t_i+t_d, E_e,c_i) G_i(E_e ) dE_e \right]
\label{liky}
\end{equation}
where of course the dependence on the 
6 model parameters is contained in $S$. 
Each detector saw $N_d$ events; 
their time, energy and cosine with supernova direction 
are  called $t_i$, $E_i$ and $c_i$ 
($i=1...N_d$). The time is counted from the first 
detected event; namely, we set $t_1\equiv 0$ for all detectors. 
The integral over the time in the first exponential 
factor is performed from the moment when the first neutrino reaches
the Earth  $t=-t_d$ (where $t_d\ge 0$), 
till the end of data taking, $t=T$ with $T=30$~s. 
The values of the 3 new parameters $t_d$, 
called `offset times',
are estimated together with model parameters by fitting the 
data (and, {\em a posteriori} found 
to be small) since the measurement of the absolute 
times in Kamiokande-II and Baksan are  not reliable. 
In IMB, the live time fraction is $f_d=0.9055$ and  
the dead time is $\tau_d=0.035$~s, whereas for the other 
detectors $f_d=1$ and $\tau_d=0.$ 
The specific background rate is $B_i=B(E_i)$ 
as discussed in \cita{prev} and in the Appendix
(we denote by $B(E_e)/2$
the background distribution, differential in time, energy
and cosine--the factor $1/2$ is for an 
uniform cosine distribution).
The Gaussian distribution
$G_i$ includes the estimated values of the energy $E_i$ and the 
error of the energy $\delta E_i$ for any individual event;
the inclusion of the error on the measurement of $\cos\theta$ 
does not add significant information, 
and the relative time of any event is 
precisely measured.

\section{\sf\color{verdon} Results of the analysis}
In this section we present the results of our statistical  analysis.
For reasons of clarity and for a more direct comparison
with previous results, we will ignore the occurrence of neutrino oscillations
in the first part of this section; 
in practice, the results that we obtain can be regarded as the best 
`effective antineutrino flux' that describes the data.
In the second part of this section, we will consider the 
modifications due to the occurrence of neutrino oscillation, and 
discuss why certain cases with inverted mass hierarchy are 
disfavored by SN1987A data.

\begin{figure}[t]
$$\includegraphics[width=0.9\textwidth,angle=0]{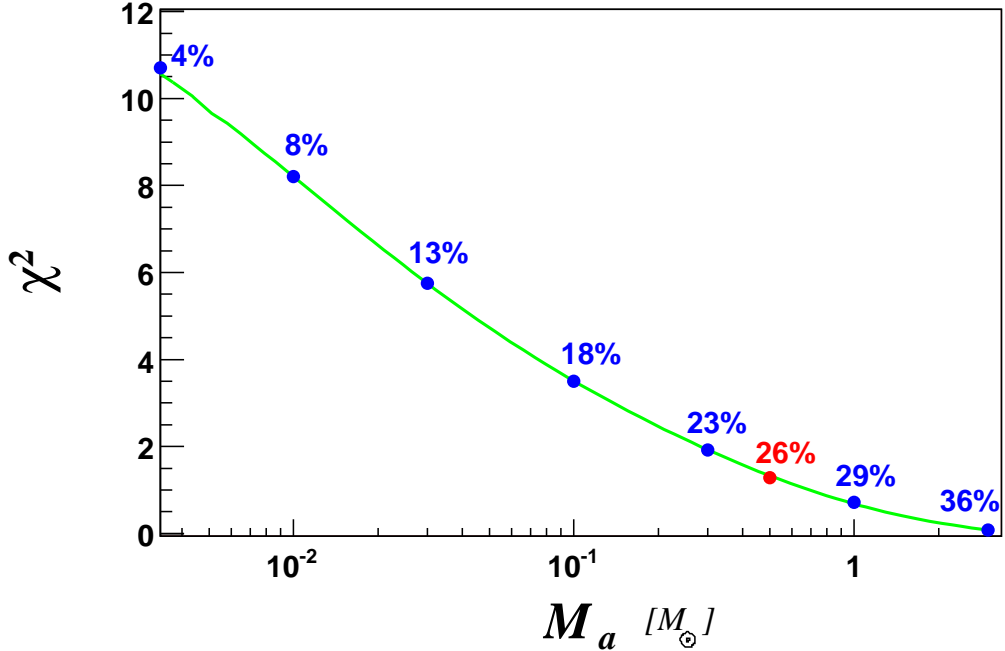}$$
\vskip-8mm
\caption{\em Values of $\chi^2$ as a function 
of the initial accreting mass $M_a$ in units of solar mass.
Also indicated the corresponding values of $f_a$ 
for the reference case $M_a=0.5 M_\odot$ and 
for the cases selected in table~\ref{tab1}. 
For $M_a=0$ (no accretion) $\chi^2=14.7$.  
\label{fig1}}
\end{figure}

\subsection{\sf\color{verdon} Fraction of energy in the accretion phase}
First of all, we searched the global minimum of the $\chi^2$.
The best fit of the initial accreting mass $M_a\sim 5\ M_\odot$ 
is not physically plausible, since we expect that the mass 
exposed to positrons will be {\em at most} the mass of the 
whole outer core $\sim 0.6 M_\odot$.
 Thus, we analyzed the $\chi^2$ as a function of $M_a$. 
For a wide range of values of $M_a$ we marginalized away  
the remaining 8 parameters by minimizing the $\chi^2$ (normalized 
as usual to the best fit point) and obtained in this 
way fig.~\ref{fig1}.
From fig.~\ref{fig1} we conclude that:\newline
$(i)$~Although larger masses would fit the 
data better, a completely reasonable value of the outer core mass,  
$M_a=0.5\ M_\odot$, is not significantly disfavored in comparison 
to the best fit value. This justifies the fact that we adopt 
this value for reference in a large part of this paper, 
rather than the best fit value.\footnote{This 
argument and this conclusion is in agreement 
with what found in \cita{ll}.}\newline
$(ii)$ There is a 
significant hint for an accretion phase.  In fact, let us ask whether 
the improvement going from $M_a=0$ (no accretion) 
to $M_a=0.5\ M_\odot$ is simply due to the presence of 
the two new variables $T_a$ and $\tau_a$ ({\em i.e.}, let us    
perform a likelihood ratio test with two degrees of freedom). 
We find that we can reject the null hypothesis in 
favor of the hypothesis that accretion occurred with a significance of 
$\alpha=1.2\times 10^{-3}$, {\em i.e.}, $3.2\sigma$ in Gaussian language.\newline 
$(iii)$~Even small values of $M_a$ are able to improve 
significantly the fit to SN1987A data. In fact, 
when $M_a$ decreases, the best fit value of the 
temperature $T_a$ increases, as one can read  
clearly from table~\ref{tab1}.
Keeping in mind eq.~\ref{fluso}, one understands 
that the number of events during accretion 
scales as $M_a\times T_a^{6-8}$, so that small changes of $T_a$ are
able to `compensate' the decrease in $M_a$.
A similar argument applies also to the values of the 
accretion energy ${\cal E}_a$ and of the energy fraction 
$f_a$, that, as we can see from table~\ref{tab1} and fig.~\ref{fig1}, 
react slowly to important changes of $M_a$.\newline 
$(iv)$ Finally, a value smaller than $M_a=0.01 M_\odot$ 
is disfavored at about 99\% in comparison to the one we selected.

Now we consider the outcome of the fit 
in the point $M_a=0.5\ M_\odot$. 
The best fit of the offset times is zero.
Their 1 sided, $1\sigma$ errors (obtained by integrating the 
marginalized likelihood) are:
\begin{equation}
\Delta t_{\rm KII}=0.09\mbox{ s}, \ \Delta t_{\rm IMB}=0.31\mbox{ s}, \ 
\Delta t_{\rm Baksan}=0.25\mbox{ s}
\end{equation}
similar results remain valid also with oscillations.
The values of the 6 astrophysical parameters 
that we find from our statistical
analysis along with the 1$\sigma$ errors 
(obtained instead by 
a conventional, $\Delta \chi^2=1$, Gaussian procedure) are: 
\begin{equation}
\begin{array}{lll}
M_a\equiv 0.5\ M_\odot, & 
T_a=2.0 \pm 0.1 \mbox{ MeV}, &
\tau_a=0.70^{+0.17}_{-0.21}\mbox{ s} \\[1ex]
R_c= 12^{+6}_{-4} \mbox{ km},& 
T_c=5.5 \pm 0.8  \mbox{ MeV}, &
\tau_c=4.4^{+1.5}_{-1.0}\mbox{ s} 
\end{array} \label{fito1}
\end{equation}
The temperature of 
the electrons $T_a$ is pretty low; this outcome of the fit is simply due to 
the fact that the early Kamiokande-II events 
have low energy \cita{prev}. The duration of the accretion 
phase is rather close to the expectations, about 
half a second, \cita{del1,del2} and \cita{totani}.
Coming to the parameters of the cooling phase, we see that the 
radius of neutrino-sphere is rather similar to the radius of the
neutron star, as expected (see {\em e.g.}, \cita{vulcano}). 
The temperature $T_c$ implies
an initial average energy $3.15 T_c$ and an average value 3/4 lower,
namely, $12.9\pm 1.9$~MeV, 
which compares well with the expectations~\cita{bahcall,keil}.
Finally, the duration of the cooling phase 
is brief, {\em e.g.}, when compared with the 20 seconds 
estimated in \cita{nad} though this leads to a reasonable value 
of the total emitted energy $2.4\times 10^{53}$~erg, that compares
well with the one expected for neutron 
star formation \cita{ns}.\footnote{The total number of 
events {\it seen / \bf expected} in 30 seconds 
(in brackets the background events) is:
\centerline{$
N_{\rm KII}={\it 16~/~\bf 20.0}\ (5.6), \ N_{\rm IMB}={\it 8~/~\bf 6.0}\ (0.0), \ N_{\rm Baksan}={\it 5~/\bf ~2.5}\ (1.0) 
$}} 
\begin{table}
\begin{center}
\begin{tabular}{|ccc|c|ccc|c|}
\hline
$M_a$ & $T_a$ & $\tau_a$ & ${\cal E}_a$ & 
$R_c$ & $T_c$ & $\tau_c$ &\ \ \ $f_a$\ \ \ \ \\[0ex]
[$M_\odot$] & [MeV] & [s] & [$10^{52}$ erg] & 
[km] & [MeV] & [s] & [\%]  \\
\hline\hline
0.003 & 3.7 & 0.69 & {\bf 1.2} & 
19 & 4.6 & 4.7 & {\bf 4}  \\ 
0.01 & 3.3 & 0.69 & {\bf 2.0} & 
15 & 4.9 & 4.8 & {\bf 8}  \\ 
0.03 & 2.9 & 0.69 & {\bf 2.9} & 
13 & 5.1 & 4.7 & {\bf 13}  \\ 
0.1 & 2.5 & 0.69 & {\bf 4.0} & 
12 & 5.3 & 4.6 & {\bf 18}  \\ 
0.3 & 2.2 & 0.70 & {\bf 5.4} & 
12 & 5.4 & 4.4 & {\bf 23}  \\ 
1.0 & 1.9 & 0.71 & {\bf 7.6} & 
12 & 5.5 & 4.3 & {\bf 30}  \\ 
3.0 & 1.6 & 0.72 & {\bf 10.} & 
12 & 5.5 & 4.1 & {\bf 36}  \\ 
\hline
\end{tabular}
\end{center}
\caption{\em The astrophysical parameters of neutrino emission 
defined in eqs.~\ref{fluso} and~\ref{tamp}, calculated for selected values of 
$M_a$. In boldface the derived quantities defined in eqs.~\ref{fafa} and~\ref{teor}.\label{tab1}}
\end{table}

{}With the parameters in eq.~\ref{fito1}
it is straightforward to calculate the amount of energy emitted 
during accretion, during cooling, and the energy fraction
$f_a$:
\begin{equation}
\left\{
\begin{array}{r}
{\cal E}_a=6.1\times 10^{52}\mbox{ erg} \\
{\cal E}_c=1.8\times 10^{53}\mbox{ erg} 
\end{array} 
\right. 
\ \Rightarrow \ f_a(\mbox{no osc.})=26\%
\label{fa-no}
\end{equation}
It is interesting to note 
that the central value of $f_a$ estimated from the fit is two-three 
times larger than the expected one \cita{10percent}
(see also \cita{coop}). 
Other values of $f_a$ are considered in table~\ref{tab1} and 
correspond to the points marked in figure~\ref{fig1}.

\subsection{\sf\color{verdon} Impact of neutrino oscillations}
We will assume that the temperature
of the muon and tau antineutrinos (that are implied in the cooling
phase) are in a fixed ratio with the $\bar{\nu}_e$ temperature.
Following \cita{keil}, we will take as default value 
\begin{equation}T(\bar\nu_\tau)/T(\bar\nu_e)=T(\bar\nu_\mu)/T(\bar\nu_e)=1.2\end{equation}
As stated above, we also assume that in 
the accretion phase muon or tau (anti) neutrinos
are absent or very rare and discuss the role of this 
assumption later.
\paragraph{\sf\color{verdon} Formalism}
Considering oscillations among the usual three neutrinos, the 
observed electron antineutrino fluxes are: 
\begin{equation}
\Phi_{\bar\nu_e}^{osc}=
P\ \Phi_{\bar\nu_e} + (1-P)\ \Phi_{\bar\nu_\mu}\label{wish}
\end{equation}
where it is assumed that $\Phi_{\bar\nu_\mu}=\Phi_{\bar\nu_\tau}$.
The expression of the electron neutrino survival probability $P$, 
that keeps into account the matter effect \cita{dighe,fogli}, is:
\begin{equation}
P=\left\{ 
\begin{array}{ll}
U_{e1}^2 & \mbox{for {\bf normal} mass hierarchy}\\[1ex]
U_{e1}^2 P_f + U_{e3}^2 (1-P_f) & \mbox{for {\bf inverted} mass hierarchy}
\end{array}
\right.\label{pale}
\end{equation}
where we have to distinguish the two 
arrangement of the neutrino mass spectrum 
compatible with present knowledge of neutrino oscillations
(see {\em e.g.} \cita{revia}).  
We adopt the conventional decomposition of the mixing elements 
in terms of the mixing  angles: 
$U_{e3}=\sin\theta_{13}$ and $U_{e1}=\cos\theta_{12}\cos\theta_{13}$.
We see that in the case of normal mass hierarchy, the probability 
$P\sim 0.7$ is reliably predicted and rather precisely known. 
Instead, for inverted mass hierarchy, 
$P$ depends strongly on the unknown mixing angle $\theta_{13}$. 
In fact, the so called flip probability $P_f$ 
(that quantifies the loss of adiabaticity
at the `resonance' related to the atmospheric $\Delta m^2$) is: 
\begin{equation}
P_f(E_\nu,\theta_{13})=\exp\left[-\frac{U_{e3}^2}{3.5\times 10^{-5}} 
\times 
\left( \frac{20\mbox{ MeV}}{E_\nu}\right)^{2/3}  \right]
\end{equation}
where the numerical value corresponds to the supernova profile 
$N_e\sim 1/r^3$ given in \cita{fogli}. For the measured solar 
oscillation parameters, the Earth matter effect is expected to be 
pretty small~\cita{cavanna}, and we will neglect it in the rest of the 
analysis. 

\paragraph{\sf\color{verdon} Results for normal hierarchy}
A fit to the data, made in the same way as 
for eq.~\ref{fito1} but accounting for oscillations 
with normal mass hierarchy yields:
\begin{equation}
\begin{array}{lll}
M_a\equiv 0.5\ M_\odot, & 
T_a=2.1 \pm 0.1 \mbox{ MeV}, &
\tau_a=0.70^{+0.19}_{-0.20}\mbox{ s} \\[1ex]
R_c= 13^{+8}_{-5} \mbox{ km},& 
T_c=5.1^{+0.9}_{-0.7}  \mbox{ MeV}, &
\tau_c=4.4^{+1.5}_{-1.1}\mbox{ s} 
\end{array} \label{fito2}
\end{equation}
The correlation coefficients $\rho(x,y)=\sigma(x,y)/(\sigma(x)\sigma(y))$, 
in percent (\%), are:\\[1ex]
\centerline{\footnotesize 
\begin{tabular}{ccccc}
$\rho(T_c,R_c)=-89$, & $\rho(T_c,\tau_c)=-42$, & $\rho(T_a,\tau_a)=-16$, & $\rho(T_a,\tau_c)=11$, & $\rho(\tau_c,R_c)=9$, \\[0ex]
$\rho(R_c,T_a)=-9$, & $\rho(\tau_c,\tau_a)=4$, & $\rho(T_c,\tau_a)=-3$, & $\rho(T_c,T_a)=1$, & $\rho(R_c,\tau_a)=0$. \\[1ex]
\end{tabular}}
The larger ones correlate the parameters of the cooling 
phase (a well-known result). 
The coefficients that correlate  the phases of cooling 
and accretion are instead small.

The $\chi^2$ increases by $\Delta\chi^2=0.8$, 
namely, it does not change significantly. In other words,
we claim that in the context of the discussion {\em SN1987A data
alone do not provide us with a strong sensitivity to oscillations
with normal mass hierarchy}. 
This can be understood as follows. 
The main impact of oscillation 
is on the neutrinos emitted during accretion;
in fact, the flux of antineutrinos  is multiplied by 
$P\sim 0.7$. This means that the expected number of events during 
accretion
is the same that we found without oscillations, for a value of 
$M_a= P\times 0.5\sim 0.35$. 
{}From table~\ref{tab1} and figure \ref{fig1} we see that 
this case has a perfectly acceptable $\chi^2$, that concludes
the explanation. However, the 
fraction of energy emitted during accretion does not diminish with 
the oscillations,  because even if 30\% of the antineutrinos 
become invisible to inverse beta decay reaction, they should be 
anyway produced. In fact, the higher temperature $T_a$ implies that 
$f_a$ increases:
\begin{equation}
\left\{
\begin{array}{r}
{\cal E}_a=8.1\times 10^{52}\mbox{ erg} \\
{\cal E}_c=1.7\times 10^{53}\mbox{ erg} 
\end{array} 
\right. 
\ \Rightarrow  \ f_a(\mbox{osc., n.h.})=32\%
\label{fa-nh}
\end{equation}
compare with 
eq.~\ref{fa-no}. Eq.~\ref{fa-no} is the main result 
of our analysis, and will be discussed in detail later.

\begin{figure}[t]
$$\includegraphics[width=0.9\textwidth,angle=0]{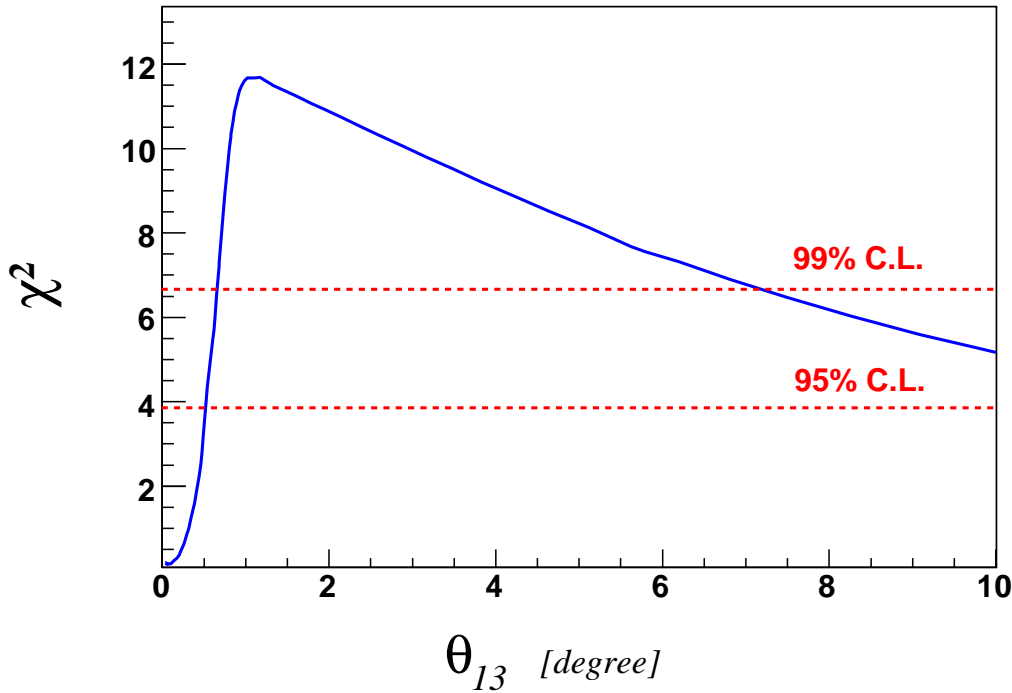}$$
\vskip-8mm
\caption{\em The $\chi^2$ for the hypothesis of inverted mass 
hierarchy and for values of $\theta_{13}$ allowed by present knowledge
of oscillations. See text for discussion and warnings.\label{fig2}}
\end{figure}

\paragraph{\sf\color{verdon} Results for inverted hierarchy}
The case of inverted mass hierarchy can produce 
$P\sim 0$ and thus a strong suppression of the 
electron antineutrino flux during accretion. 
This can be seen clearly from figure \ref{fig2},
where (fixing $M_a=0.5 M_\odot$) we give the value of  
the $\chi^2$ for various values of $\theta_{13}$ that are allowed
by what we know at present on oscillations. 
Thus, the  conventional treatment of oscillations 
in eq.~\ref{pale} implies that
the high-luminosity feature visible in the first second of 
emission disfavors 
the case when neutrinos have an 
inverted mass hierarchy and values of 
$\theta_{13}$ larger than $0.5^\circ-1^\circ$.
It should be noted that for the largest 
value of $\theta_{13}$ the fit improves, but 
the price to pay is a very large amount of energy 
emitted during accretion; {\em e.g.}, for $\theta_{13}=10^\circ$
we need ${\cal E}_a=7.7\times 10^{53}$~erg $>{\cal E}_c=2.1\times 10^{53}$~erg!
This can be seen as an indication against the case of inverse mass hierarchy 
and relatively large values of $\theta_{13}$. 
This result is surely interesting, 
but in our opinion it should be 
taken with caution due to:\\
\noindent 1.~Limited astrophysical information. Indeed, 
in presence of a relevant muon/tau antineutrino emission during accretion,
the effect of oscillation weakens (at the same time the fraction $f_a$
increases, see eq.~\ref{teor} and discussion therein).\\
\noindent 2.~Possible modifications of physics of oscillation. Indeed,
neutrino contribute to weak (`matter') potential, especially during 
accretion, making the problem non-linear \cita{fuller}.\\
Our treatment of oscillation conforms to the conventionally
accepted framework and develops in the context of eq.~\ref{teor}.
This is the best that we can do at present;  
once the modifications from points 1.~and 2.~above 
will be precisely quantified, it will be a straightforward exercise to repeat
these steps to know whether these results change significantly.

However, it is interesting to try to explore 
what could happen by relaxing the assumptions
we followed till now, and this is what we attempt 
in the last part of this section. 
For definiteness, 
let us ask what happens if 
$P=\kappa$ with $\kappa=0.1$ or $=0.5$.
The  fit is similar to the one corresponding to 
$M_a=P\times 0.5 M_\odot$ in table~\ref{tab1}
(thus acceptable on the basis of $\chi^2$)
however we would need to emit during accretion an energy 
$1/P$ times larger than the one we read from table~\ref{tab1}.
More in details, we find
${\cal E}_a= 3.3\times 10^{53}$~erg for $\kappa=0.1$ 
and ${\cal E}_a= 1.0\times 10^{53}$~erg for $\kappa=0.5$.
The first case can be questioned on theoretical basis, whereas
the second, that leads to $f_a=36$\%, 
seems much more reasonable.  
This means that if non-linear effect are able to produce 
an effective survival probability  $P\sim 0.5$, 
inverted neutrino oscillations will be not in 
serious disagreement with SN1987A observations.
Comparing with eq.~\ref{wish}, 
we see that the two case here considered are similar 
to the case $P\sim 0$ ({\em i.e.}, oscillations conforming to the 
conventional expectations) but 
with a flux of non-electronic 
neutrinos  
during accretion $\Phi_{\bar\nu_x}=\kappa \times \Phi_{\bar\nu_e}$.
However, the estimated values of ${\cal E}_a$ should be increased 
by the factor $1+2 \kappa$ 
due to the presence of non-electronic species
during accretion.
We conclude that within the 
conventional theoretical framework, we can have a 
good fit of SN1987A neutrinos only with important 
modifications of the hypothesis on the 
flavor composition of the  neutrino flux during accretion (eq.~\ref{teor})
and/or with a strong modification of the oscillation probabilities
shown in eq.~\ref{pale}. These effects could save the inverted hierarchy case,
but would not change the conclusion on $f_a$, that can only increase
in comparison with the case when oscillations are neglected and presumably 
also with the case of oscillations with normal mass hierarchy.

\subsection{\sf\color{verdon} Expected range of $f_a$ and stability of 
the result}
{}From here on, we focus on the case of oscillation with normal
hierarchy; similar conclusions apply 
in the case of inverted with small $\theta_{13}<1^\circ$ 
(that becomes indistinguishable from the previous 
one for very small values of $\theta_{13}$) and also without oscillations.

We evaluated the expected range of $f_a$ in eq.~\ref{fa-nh} 
as follows:\\
\noindent $i)$ First we found the impact of $M_a$. 
By considering the $\chi^2(M_a)$ allowed by the data (as in fig.~\ref{fig1})
we obtain a corresponding range for $f_a$.
We find that $f_a>18\% \mbox{ at 95\% C.L}$.\\
$ii)$ Next, we found the effect of 
the other 5 astrophysical parameters. By propagating the errors, we 
obtain the absolute error $\delta\! f_a=10\%$.
This means that $f_a>16\% \mbox{ at 95\% C.L}$.\\
The large errors are due to the fact the the number of events 
occurring during accretion is only about~10.\footnote{This will be 
certainly much larger when we will observe the 
neutrinos from a galactic supernova.}

We also checked the stability of our results on $f_a$: \\
\noindent $a)$ By shifting the offset times within their error-bars. \\
\noindent $b)$ By considering alternative, reasonable parameterizations 
of the antineutrino flux such as those proposed and studied by \cita{ll}.\\
\noindent $c)$ By removing some event from the dataset ({\em e.g.}, 
when this is done with the event number 10 of 
Kamiokande-II--see \cita{prev}--the $\chi^2$ somewhat improves). \\
In none of these cases, however, our conclusion on $f_a$ 
does change significantly. The reason is simply that this result 
is not due to an {\em ad hoc} statistical analysis, but to two 
clear experimental facts:\\
\noindent {\em 1)} there is a relatively large number of 
events in the first second, especially in Kamiokande-II dataset;\\
\noindent {\em 2)} their energy is relatively low: compare with Sect.~3.2.3 of \cita{prev}.

In conclusion, the data suggest that a considerable  
fraction energy was emitted during accretion. 
Though it is possible to have a smaller $f_a$  
for certain values of the  
astrophysical parameters of the neutrino emission
that permit a reasonable description of the data, 
the fact that these parameters are not known 
at present and the occurrence of oscillations 
suggest that $f_a$ was 20\% or larger.

\section{\sf\color{verdon} Discussion}
We showed that data from SN1987A imply 
at 3.2$\sigma$ the existence of an 
initial phase of intense neutrino luminosity, that resembles 
what is expected during accretion. 
The agreement of the various 
astrophysical parameters with expectations is 
rather good, but there is a hint that the amount of energy 
emitted during accretion is a few times larger than what is 
expected in standard calculations.
The significance of this hint is limited by the small number 
of detected events, but at the same time, it 
is a stable outcome of the analysis and it is 
due to clear features of the data (especially,
those of Kamiokande-II).
This result is not contradicted but rather reinforced 
if three flavor oscillations of neutrinos occurred with 
normal mass hierarchy (see eq.~\ref{fa-nh} and discussion therein).
Instead, it is not easily reconciled with oscillations, 
if the mass hierarchy is inverted and $\theta_{13}$ is 
larger than about $1^\circ$. 

Let us comment  these results. 
The first implication regards the nature of the explosion.
We feel motivated to propose the speculation that 
a value $f_a\sim 20$ \% (or larger)
is a characteristic aspect of successful supernova explosions. 
It will be interesting to see whether such an 
expectation, driven by SN1987A data, will meet the findings 
of future successful simulations of supernova explosions:
see \cita{lasto} for a status report.

The second implication regards neutrino oscillations. 
We showed that a large oscillation effect on 
electron antineutrinos 
({\em i.e.}, a small survival probability $P$ in eq.~\ref{wish}) 
is disfavored by the data, since it would tend to dilute 
the number of events expected in the first second. 
This result should be taken with caution, since it is possible  
to argue that the expectations on flavor partition during accretion
are not completely reliable, and/or that the 
oscillation could have a non-conventional character \cita{fuller}.
We believe that it is urgent to clarify this issue, 
for such a result could have important implications  for 
several future experiments (such as long-baseline experiments, 
double beta decay search, cosmology~\cita{revia}).

Of course, the most important task is left 
for the future supernova neutrino 
experiments, that should study precisely the 
first second of supernova explosion: in particular, the 
number of the events and their energy.
It would be interesting to monitor the 
total flux of neutrinos in the first second, 
{\em e.g.} counting neutrons in a 
heavy water detector as SNO by $\nu D\to \nu p n$. 
However, it seems fair to conclude that 
the traditional method of investigation
(namely, the observation of supernova electron antineutrino
by inverse beta decay in scintillators or water \v{C}erenkov 
detectors, possibly allowing for improvements in 
neutron detection) has still a bright future.

In conclusion, we presented a 
state-of-art analysis of the first second of 
SN1987A neutrino emission. We hope that this
can be useful to understand better 
what happened in this crucial moment 
of neutrino astronomy and, possibly, to 
make further steps toward the theory of 
supernova explosion.

\subsection*{\sf\color{verdon} Acknowledgment}
We thank D. K. Nadyozhin for discussions. 
Preliminary results were presented by G.P. in \cita{giulia}.

\appendix\section{\sf\color{verdon} Peculiarities of the analysis\label{sec:app}}
As recalled, our analysis is similar 
to the one of Lamb and Loredo \cita{ll}, 
with whom we agree within errors when we 
strictly stick to their procedure. 
The analysis that we eventually adopted in this paper  
departs from their one for the inclusion of oscillations,
and for some technical points that we describe here:

\noindent
{\em 1) Background:} Lamb and Loredo 
fold the {\em measured} background curve 
with the distribution of the energy 
distribution of the events. However, this has the effect of 
double counting the detector-dependent effects on energy measurement, 
for which the detected energy is distributed around the true energy
(`smearing'). Thus, we prefer to directly use the background curve,
setting $B_i=B(E_i)$; namely, we do not perform any folding. 
This has the consequence that
the events of Kamiokande below 7 MeV have a higher background rate,
and those above 9 MeV a lower background rate, while the other ones 
stay almost unchanged. The changes for Baksan are instead negligible.
See \cita{prev}, Appendix~A.

\noindent
{\em 2) Cross section:} When evaluating 
the signal rate $S$ (eq.~\ref{seta}) we adopt the 
inverse beta decay cross section
calculated in \cita{sss}. We use the expression for the 
cross section $d\sigma/d\cos\theta$ given in eq.~(20) there.
The energy of antineutrino is given in term of the 
positron energy $E_e$ and the angle $\theta$ 
between the antineutrino and the positron directions:
\begin{equation}
E_\nu=\frac{E_e+\delta}{1-(E_e-p_e\cos\theta)/{m_p}},
\end{equation}
where $\delta=(m_n^2-m_p^2-m_e^2)/(2 m_p)\approx
1.294\mbox{ MeV}$. 
We replace $\cos\theta$ in the previous equation with the measured
values for the first 12 Kamiokande-II and for the 8 IMB events, and 
set instead $\cos\theta=0$ for 
the 5 events of Baksan and the last 4 events 
of Kamiokande-II.

\noindent
{\em 3) Efficiency:} 
Following \cita{ll}, we average the signal $S$ 
as a function of the true value of the energy $E_e$ 
over its distribution (assumed to be Gaussian); 
namely, we keep into account the energy 
smearing of the signal: see eq.~\ref{liky}. 
But differently from \cita{ll}, we include also the detection 
efficiency as a function of the true energy of the event:
see eq.~\ref{seta}. With this procedure, we describe 
the fact that the expected numbers of signal events 
({\em i.e.,} the crucial input of the likelihood)
should include all relevant detector 
dependent features (such as loss of events 
due to light attenuation, fluctuations of the number of 
photoelectrons, detector geometry, etc) including 
those that lead to an imperfect ($\eta_d<1$) 
detection efficiency.\footnote{We 
have in mind an `average efficiency'  evaluated by a MC procedure, 
namely 1)~by  simulating several events with {\em true}
energy of the positron $E_e$ but located in the various 
positions and emitted in all possible directions, 
and then 2)~counting the fraction of times that an event is recorded 
and finally 3)~deducing also the smearing on the energy 
(=average error as a function of $E_e$).
For an even more refined analysis of SN1987A data, 
one should evaluate for any individual event 
the specific detection efficiency and 
background rate \cita{ll,prev}. In our understanding,
such a correction on individual basis was performed only to
assess the errors on the energies of the events, see \cita{kam}.} 
We argue in favor of our procedure by considering the following situation. 
Imagine two detectors
where the signal interaction rate is equal to the background rate,
that differ by the detector efficiency:
$\eta=100\%$ in the first one, $\eta=10\%$ in the second.
Now, suppose that each of them observes an event.
According to the procedure in \cita{ll}, 
the probability that the observed event is due to a signal 
is 50\% in both detectors (see table IV and discussion therein), 
that we find paradoxical. Instead, adopting our procedure the probability 
that the event is due to signal is 
50\% in the first detector and 9\% in the second one, which seems 
to us more reasonable.


\noindent
{\em Impact of the modifications:}
The most important effect is the inclusion of the efficiency of detection,
followed by the new cross section and finally by our assumption on the 
background. 
{\em E.g.}, adopting the procedure of \cita{ll}
the temperature parameter $T_c$, the radius $R_c$ and 
the time constant $\tau_c$  
of an exponential cooling model (as in eq.~\ref{fluso} 
but with $M_a=0$)
are 3.7~MeV, 44 km and 
4.4 s. When we include the efficiency, 
they become 4.2~MeV, 30 km and 3.9~s.
When we use also the new cross section
(eq.~(25) in~\cita{sss}) they become 4.6~MeV, 26 km and 3.7 s,
and when we include the dependence 
on the $\cos\theta$  (eq.~(20) in \cita{sss}),
these values become 4.5~MeV, 27~km and 3.8 s.
Finally, with the new background all quantities
stay practically unchanged.

\footnotesize

\begin{twocolumn}

\section*{\sf\color{verdon}  References}
\def\refname{

\end{twocolumn}

\end{document}